\documentclass[conference]{IEEEtran}
\IEEEoverridecommandlockouts
\usepackage{cite}
\usepackage{amsmath,amssymb,amsfonts}
\usepackage{algorithmic}
\usepackage{graphicx}
\usepackage{hyperref}
\usepackage{textcomp}
\usepackage{xcolor}
\def\BibTeX{{\rm B\kern-.05em{\sc i\kern-.025em b}\kern-.08em
    T\kern-.1667em\lower.7ex\hbox{E}\kern-.125emX}}
\begin{document}

\title{Resource Constrained U-Net for Extraction of Retinal Vascular Trees\\
}
\makeatletter
\newcommand{\linebreakand}{%
  \end{@IEEEauthorhalign}
  \hfill\mbox{}\par
  \mbox{}\hfill\begin{@IEEEauthorhalign}
}
\makeatother

\author{\IEEEauthorblockN{1\textsuperscript{st} Georgiy Kiselev}
\IEEEauthorblockA{\textit{Department of Statistics} \\
\textit{University of California, Davis}\\
Davis, USA \\
gakiselev@ucdavis.edu}
}

\maketitle

\begin{abstract}
This paper demonstrates the efficacy of a modified U-Net structure for the extraction of vascular tree masks for human fundus photographs. On limited compute resources and training data, the proposed model only slightly underperforms when compared to state of the art methods. 
\end{abstract}


\section{Introduction}
\IEEEPARstart{T}{he} analysis and diagnosis of retinal vascular structures play a crucial role in the early detection and management of various ocular and systemic diseases [1], including diabetic retinopathy, hypertension, and cardiovascular conditions. Retinal imaging, particularly fundus photography, provides a non-invasive method to capture detailed views of the retina, facilitating the assessment of vascular health. However, the manual segmentation and analysis of retinal vessels are labor-intensive, time-consuming, and prone to inter-observer variability. This has led to a significant interest in automated techniques for retinal vessel extraction, which promise to enhance diagnostic accuracy and efficiency.

\section{Background}
Prior advancements in deep learning, particularly convolutional neural networks (CNNs), have revolutionized the field of medical image analysis. Among these, the U-Net architecture proposed by Ronneberger et al. [2] has emerged as a highly effective model for biomedical image segmentation due to its ability to learn from a relatively small amount of annotated data and its robustness in capturing fine-grained structures. The U-Net's encoder-decoder structure, coupled with its skip connections, allows for the precise localization of features and has been successfully applied to various segmentation tasks in medical imaging.

Despite these advances, the application of U-Net specifically for the extraction of retinal vascular trees presents unique challenges. The retinal vasculature consists of thin, elongated structures with varying contrasts and overlapping regions, making it difficult for standard segmentation approaches to accurately delineate the vessels from the background. Previous research has demonstrated the potential of U-Net in this domain, but there remains a need for further optimization and validation to ensure its reliability and generalizability across diverse datasets.

In this paper, we propose an application of the standard U-Net architecture with added Batch Normalization and Dropout layers for the extraction of retinal vascular trees. Our research is motivated by the need to improve the accuracy and efficiency of retinal vessel segmentation, which is essential for advancing automated retinal diagnostics and facilitating large-scale epidemiological studies. We aim to address the following key question: How can the U-Net architecture be optimized to achieve precise and reliable segmentation of retinal vascular structures across different imaging conditions and datasets?
\section{Literature Review}
The application of machine learning techniques towards the extraction of retinal vessels is by no means a novel approach. However, there have been several different methods employed that affect how important information is extracted. 

Supervised classification techniques have long been used in retinal vessel segmentation. A seminal study by Staal et al. [3] explored supervised classification methods, emphasizing the importance of feature extraction and classifier selection in achieving high segmentation accuracy . This research laid the groundwork for subsequent studies that combine traditional machine learning approaches with deep learning methods to improve segmentation performance.

Trainable COSFIRE (Combination of Shifted Filter Responses) filters represent another significant advancement in retinal vessel segmentation. Azzopardi et al. [4] proposed this method, which involves training filters to detect specific patterns, such as blood vessels, by combining responses from multiple shifted versions of the filter. This approach has shown considerable promise in enhancing the accuracy of vessel detection, particularly in challenging imaging conditions.

In the realm of automatic blood vessel segmentation, Osareh et al. [5] and Roychowdhury et al. [6]  presented a comprehensive method that integrates preprocessing, vessel enhancement, and segmentation into a cohesive framework. This study highlighted the potential of combining multiple stages of image processing to improve the overall performance of vessel segmentation algorithms.

Cross-modality approaches, which utilize data from different imaging modalities, have been investigated to enhance retinal vessel segmentation. Qiaoling et al. [7] explored a method that integrates information from multiple imaging sources to improve segmentation accuracy and robustness. This technique underscores the potential of combining complementary data to achieve more comprehensive segmentation results. Ensemble-based methods, which combine the predictions of multiple classifiers, have also been explored for retinal vessel segmentation. Fraz et al. [8] demonstrated the effectiveness of ensemble learning in improving segmentation accuracy by leveraging the strengths of different classifiers. This approach has proven particularly useful in reducing the variance and bias associated with individual classifiers.

The application of deep neural networks (DNNs) in vessel segmentation has seen significant advancements. Zhang et al. [9] proposed a DNN-based method for retinal vessel segmentation, demonstrating its superior performance compared to traditional methods. Similarly, Liskowski [10] highlighted the benefits of using deep learning models, such as convolutional neural networks (CNNs), for accurate and efficient vessel segmentation.

Our proposed method introduces Batch Normalization and Dropout layers into the original U-Net structure, which should help solve overfitting and performance issues faced by a base U-Net.

\section{Data}
\subsection{Description}
For this paper, we utilize the Digital Retinal Images for Vessel Extraction (DRIVE) dataset. It was made freely available online as part of a diabetic retinopathy screening program in the Netherlands. The dataset consists of 40 high resolution images of human retina and corresponding manually annotated binary masks of their vascular trees. All images are 584 pixels tall and 565 pixels wide with 3 color channels. The corresponding ground truth masks are the same height and width, with one color channel (binary masks). In this paper, training inputs and predicted masks will be of size 512 by 512 pixels in order to simplify calculations within the network. 

\subsection{Exploratory Analysis}
To understand the visual quality and consistency of the annotations, we randomly selected a subset of image-mask pairs from the dataset. These samples highlight the complexity and variability of retinal vasculature, as well as the accuracy and granularity of the manual annotations. The masks clearly delineate the vessel structures, which will be the target for our UNET segmentation model (Fig. 1). 
\begin{figure}[htbp]
\centerline{\includegraphics[width=90mm,scale=0.7]{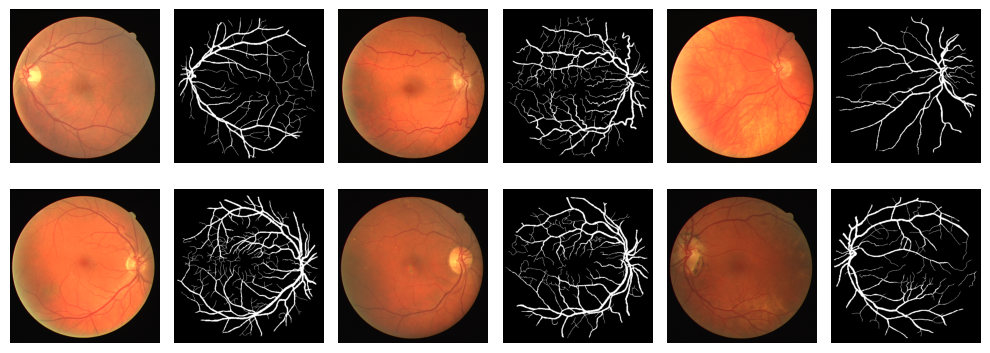}}
\caption{Original image-mask pairs.}
\label{fig}
\end{figure}

Calculating the mean image and mask provides insights into the average characteristics of the dataset. The mean image (Fig. 2a) reveals the general brightness and common features across all retinal images, while the mean mask (Fig. 2b) shows the typical distribution of vessel locations and their density.

The mean image retains the circular structure of the retina with a high-intensity region corresponding to the optic disc, and a dark, low intensity region corresponding to the fovea. The mean mask, on the other hand, accentuates the primary vessel pathways, emphasizing areas of higher vessel density typically observed in retinal images.

The figure also demonstrates a good property - each person's retinal vascular tree is unique, which could inspire further research on the use of vessel extraction for biometric identification. 
\begin{figure}[htbp]
\centerline{\includegraphics[width=90mm,scale=0.7]{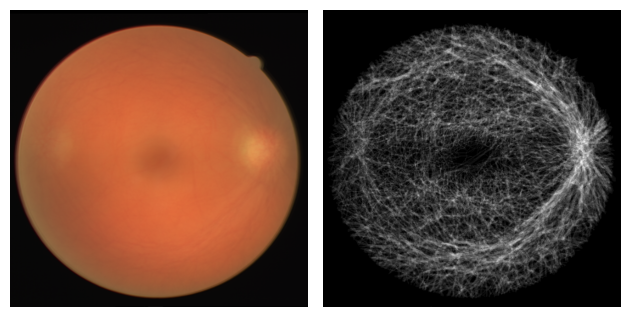}}
\caption{Mean image-mask pair}
\label{fig}
\end{figure}

Analyzing the pixel value distribution of the images offers further understanding of the dataset's intensity characteristics. The histogram of pixel intensities (Fig. 3) for retinal images displays a multimodal distribution, which is typical for such data due to the contrast between vessels and the background.

The histogram shows a significant peak at the lower end of the intensity spectrum, corresponding to the dark background of the retina, and several smaller peaks at higher intensity values, corresponding to the blood vessels and optic disc. The final significant peak at pixel value 255, (corresponding to the color white) represents the intensity seen at the optic disk we see in Fig. 2a.

\begin{figure}[htbp]
\centerline{\includegraphics[width=90mm,scale=0.7]{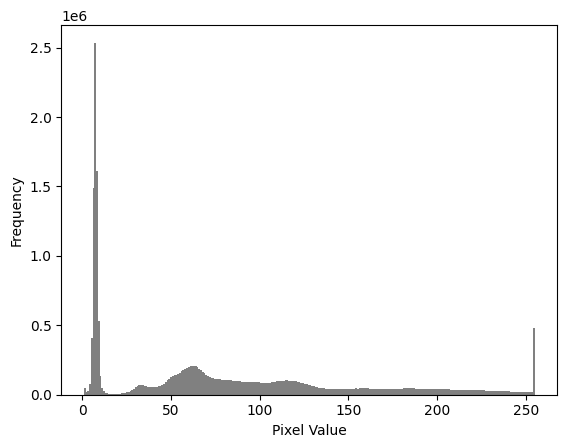}}
\caption{Image pixel value distribution}
\label{fig}
\end{figure}

To explore the relationship between pixel values in the images, we plotted the distribution of correlation coefficients between corresponding pixels (Fig. 4). This histogram illustrates the correlation across each possible image-image pair, indicating how consistently pixel values in the images correspond to one another. We disregard the bar at correlation coefficient 1, because that is the correlation of an image with itself.

The distribution shows that most correlation coefficients are above 0.9, suggesting a strong positive correlation between image pixel intensities in the data. This consistency is crucial for training effective segmentation models, as it indicates that pixel intensities reliably represent vessel structures and do so consistently across all images.

\begin{figure}[htbp]
\centerline{\includegraphics[width=90mm,scale=0.7]{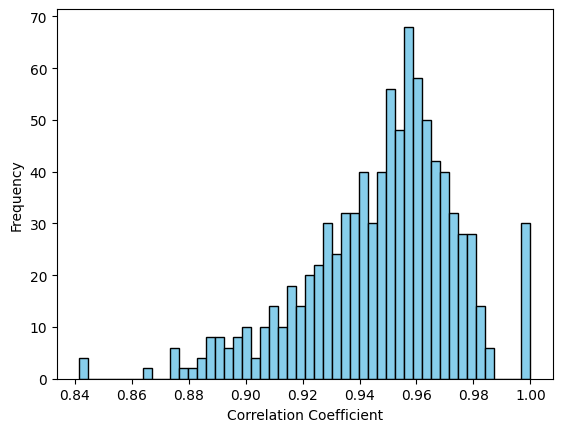}}
\caption{Correlation between training image pixel values}
\label{fig}
\end{figure}

\subsection{Preliminary Augmentations and Further Analysis}
To enhance the contrast and visibility of retinal structures, we converted the original color images to grayscale and normalized them. This transformation allows for more effective extraction of features such as blood vessels and lesions. Figure 5 compares the original color image with its grayscale counterpart, highlighting the improved contrast of vascular structures in the grayscale image. Because we are generating a binary mask with one color channel as the output, we deem that the color of the image is unimportant for the model to train on. 

\begin{figure}[htbp]
\centerline{\includegraphics[width=90mm,scale=0.7]{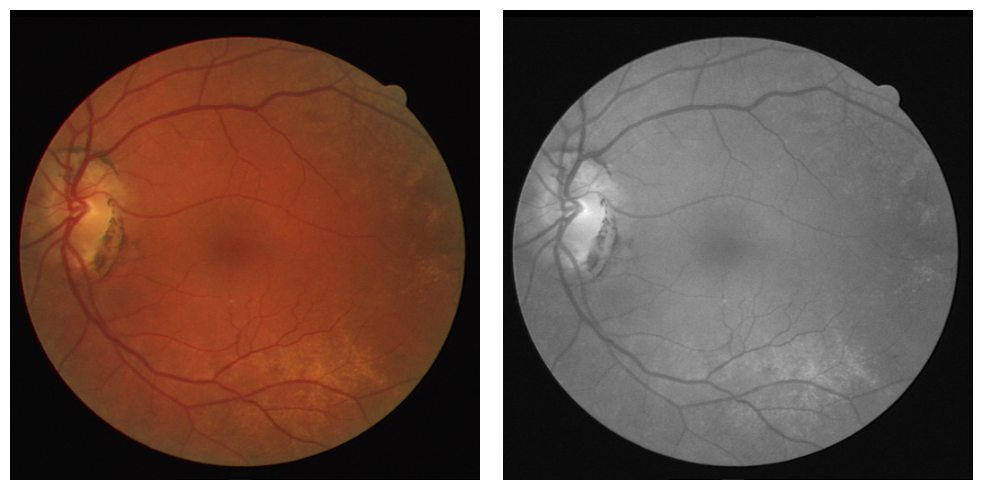}}
\caption{Original and grayscale image.}
\label{fig}
\end{figure}

We then applied Contrast Limited Adaptive Histogram Equalization (CLAHE) to the grayscale images. CLAHE is an advanced histogram equalization technique that improves local contrast, making finer details in the retinal images more discernible. Figure 6 illustrates the effect of CLAHE on the grayscale images. The blood vessels and other anatomical features are more pronounced, facilitating better segmentation and analysis by the neural network.

\begin{figure}[htbp]
\centerline{\includegraphics[width=90mm,scale=0.7]{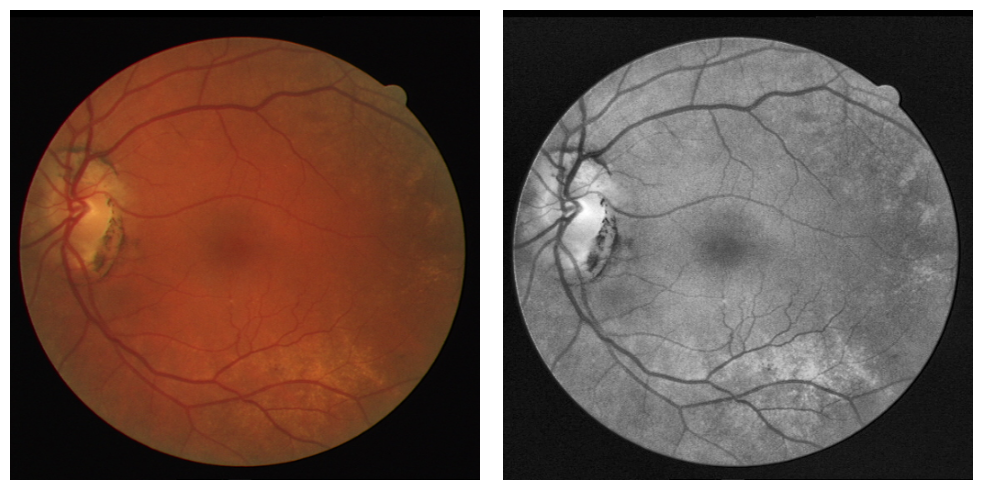}}
\caption{Original and CLAHE enhanced image.}
\label{fig}
\end{figure}

Given the limited size of the initial dataset, consisting of only 30 images, we employed data augmentation techniques to artificially increase the dataset size to 120 images. The augmentations included:

\begin{itemize}
\item \textbf{Horizontal Flip}: Images were flipped horizontally, which helps in addressing the variability in the orientation of the retinal images.
\item \textbf{Vertical Flip}: Images were flipped vertically to further enhance the diversity of the dataset.
\item \textbf{Rotation}: Random rotations were applied to the images, simulating different viewing angles and improving the robustness of the model.
\end{itemize}

Figure 7 demonstrates a random sample of augmented images from the new dataset. 

\begin{figure}[htbp]
\centerline{\includegraphics[width=90mm,scale=0.7]{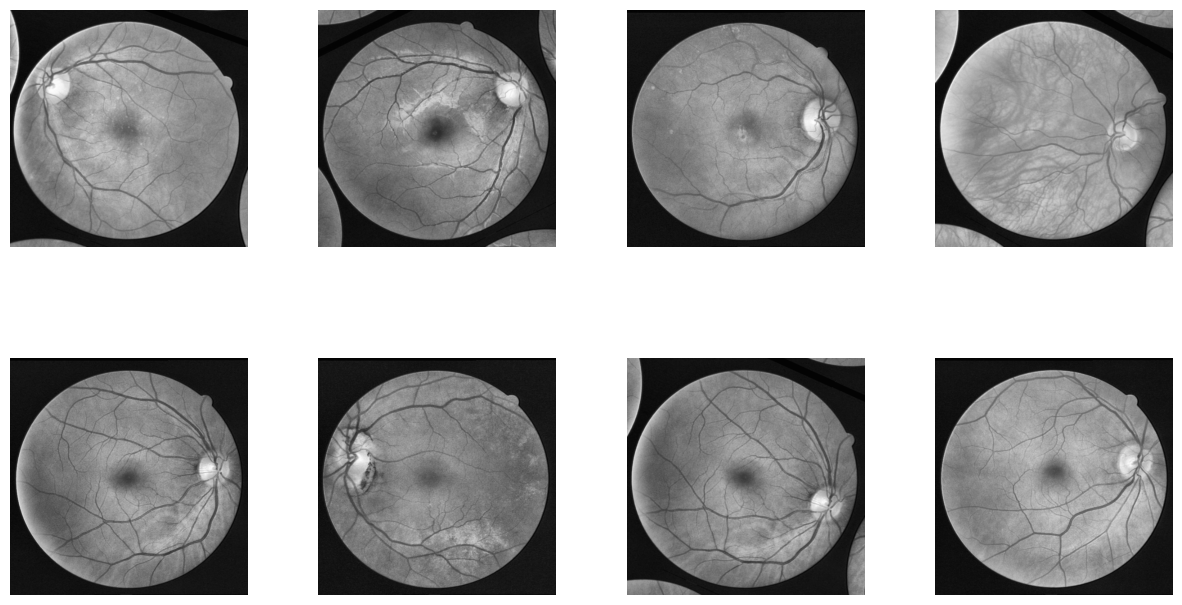}}
\caption{Random sample of augmented images.}
\label{fig}
\end{figure}

By applying these augmentations, we created additional variations of the original images, significantly enhancing the dataset's size and variability. This augmentation process ensures that the neural network is exposed to a wider range of scenarios during training, leading to improved generalization and performance on unseen data.\

\section{Methods}
\begin{figure}[htbp]
\centerline{\includegraphics[width=90mm,scale=0.7]{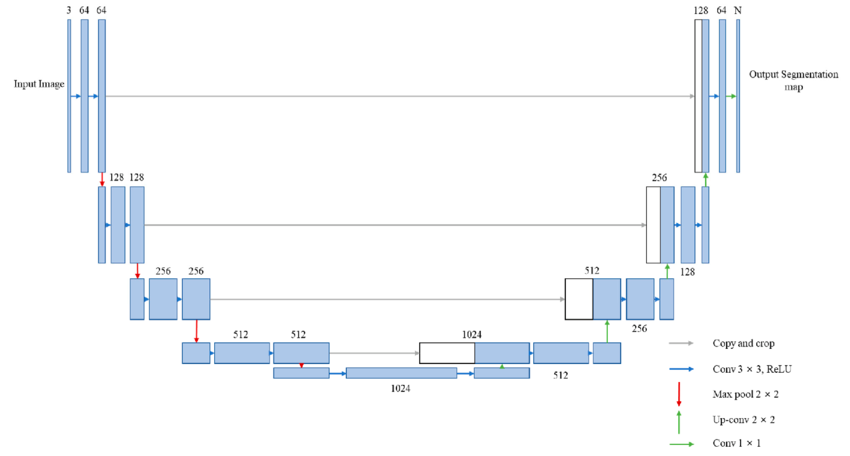}}
\caption{UNet network structure adapted with permission from Ref. [2]. Copyright 2015, copyright Ronneberger.}
\label{fig}
\end{figure}
The U-Net architecture employed in this study (Fig. 8) follows the classical design as proposed by Ronneberger et al. (2015) but includes additional batch normalization layers to enhance training stability and performance. The network architecture can be divided into two main parts: the downward path (encoder) and the upward path (decoder), with each path composed of multiple convolutional blocks.\\

The downward path consists of several convolutional blocks, each containing two convolutional layers followed by a rectified linear unit (ReLU) activation function and a max pooling layer for downsampling. The steps are detailed as follows:\\

\textbf{Input Layer:} The input to the network is an image of size H×W×C, where H and W are the height and width of the image, and C is the number of channels. \\\\
\textbf{Convolutional Block:} Each block in the downward path performs the following operations:
    \begin{itemize}
        \item \textbf{Convolution:}The 3x3 convolutional layer applies a set of filters (kernels) to the input image or feature map, producing an output that highlights specific patterns such as edges, textures, or other features. This operation preserves the spatial dimensions of the input. The mathematical operation can be represented as:
        \begin{align}
        y_{i,j,k} = \Sigma_{m=-1}^{1} \Sigma_{n=-1}^{1} \Sigma_{c=1}^{C} x_{i+m,j+n,c} \cdot w_{m,n,c,k} + b_{k}
        \end{align}
        among them, $x$ is the input, $w$ are the convolutional filters, $b$ are biases, and $y$ is the output feature map. The layer uses a ReLU activation function, which applies an element-wise non-linearity to introduce non-linearities into the model, allowing it to learn more complex functions. The ReLU function is defined as:
        \begin{align}
            y = max(0,x)
        \end{align}

        \item \textbf{Batch Normalization:} The batch normalization layer normalizes the output of the convolutional layer by adjusting and scaling the activations. The process stabilizes the learning process and accelerates convergence. It operates as follows:
        \begin{align}
        \hat{x_i} = \frac{x - \mu_{b}}{\sqrt{\sigma_{b}^2 + \epsilon}}
        \end{align}
        Here, $\mu_{b} \text{ and } \sigma_{b}^2$ are the batch mean and variance, respectively. \\ 
        \item \textbf{Convolution:} Another 3x3 convolution with ReLU activation, using the previous layer's output. \\
        \item \textbf{Batch Normalization:} A batch normalization layer is added, using the previous layer's output. \\ 
        \item \textbf{Max Pooling:}  The 2x2 max pooling operation reduces the spatial dimensions by selecting the maximum value in each 2x2 block of the feature map. This downsampling operation retains the most important features while reducing computational complexity and sensitivity to spatial translations.\\
    \end{itemize}

We then repeat the above process with the output, doubling the number of feature channels after each block, until the bottleneck of the network is reached. \\

\textbf{Bottleneck:} The bottleneck layer contains two convolutional layers with batch normalization and ReLU activations but without max pooling. This layer captures the most compressed and abstract representation of the input data. \\ 

The upward path restores the spatial dimensions while combining features from the downward path to produce a detailed segmentation map. \\

\textbf{Upsampling:} Each upsampling step involves a 2x2 transposed convolution (up-convolution), which doubles the spatial dimensions of the feature map. This operation can be seen as the inverse of the convolution operation, effectively reconstructing the image resolution:
\begin{align}
    y_{i,j,k} = \Sigma_{m=0}^{1} \Sigma_{n=0}^{1} x_{i//2+m,j//2+n,k} \cdot w_{m,n,k} + b_{k}
\end{align}

\textbf{Convolutional Block}: Similar to the downward path, there is a convolutional block following the upsampling step. The number of feature channels is halved after each block, progressively refining the feature maps to produce the final segmentation.\\

\textbf{Skip Connections}: The upsampled feature map is concatenated with the corresponding feature map from the downward path (skip connections). This step allows the decoder to leverage high-resolution features from the encoder, ensuring finer details are preserved. Skip connections help in the precise localization by combining coarse high-level information from the decoder with the fine-grained low-level information from the encoder:
\begin{align}
    y=[y_{upsampled},y_{downwards}]
\end{align}
\textbf{Output Layer}: The final layer is a 1x1 convolution that reduces the depth of the feature map to the number of desired output classes. This layer maps the high-dimensional feature vectors to the target segmentation classes:
\begin{align}
    y_{i,j,k} = \Sigma_{c=1}^{C}x_{i,j,c} \cdot w_{c,k} +b_{k}
\end{align}
A sigmoid activation is applied for binary segmentation, producing a probability map for each pixel, which we send to 0 or 1 via a boolean function with threshold 0.5. 

\subsection{Training Procedure}
The network was then trained on Google Colab's 2x T4 GPU runtime, using the Adam optimizer with an initial learning rate of $1e-4$. The loss function used was Multi-loss 'DiceBCE', combining Binary Cross Entropy and Dice losses. The training was conducted over 50 epochs with a batch size of 2 (The runtime ran out of VRAM with larger batches). 

\section{Results and Evaluation}
We evaluated the performance of our U-Net model for retinal vessel segmentation using a comprehensive set of metrics and visualizations. The evaluation metrics include the Jaccard Index, F1 Score, Recall, Precision, Accuracy, and ROC AUC. Additionally, we provide insights from the training and validation loss curves.
\subsection{Predicted Masks}
In this subsection, we analyze the predicted masks generated by our U-Net model for retinal vessel segmentation. A representative example of a predicted mask alongside the corresponding ground truth mask is presented in Figure 9. This visual comparison highlights the model's ability to segment retinal vessels with notable accuracy while also revealing areas where the model's performance could be improved.
\begin{figure}[htbp]
\centerline{\includegraphics[width=90mm,scale=0.7]{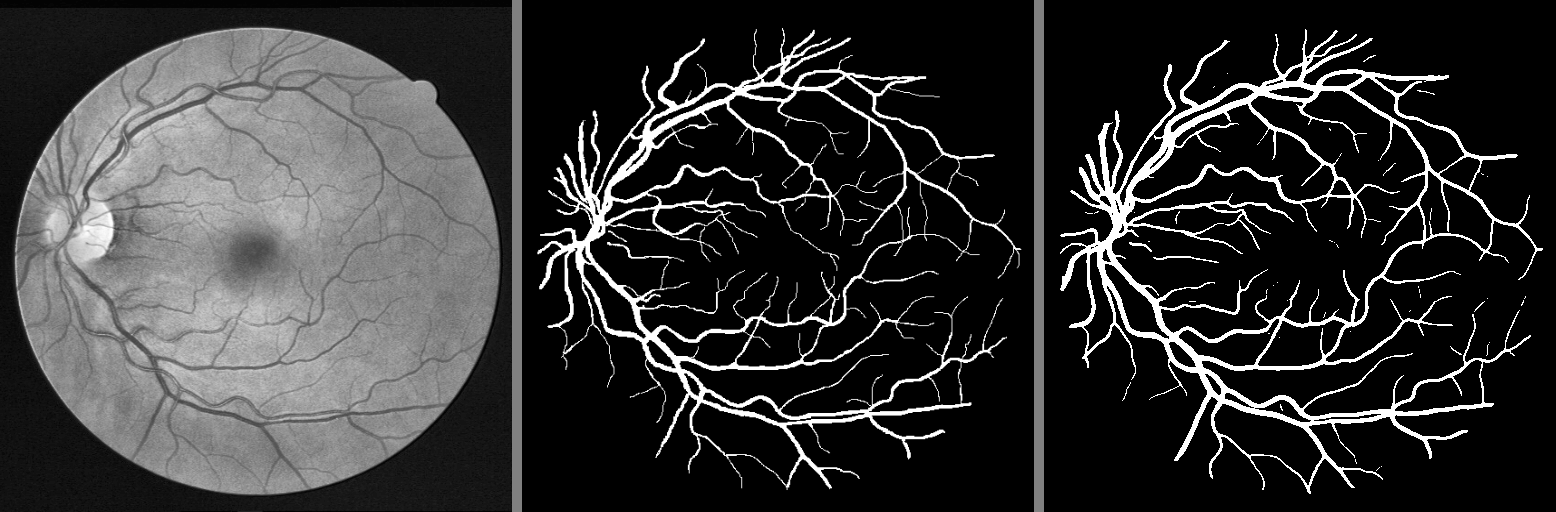}}
\caption{Test image, true and predicted masks.\textsuperscript{a=}}
  \small\textsuperscript{a=} Note: Test images are modified via CLAHE as part of predicted mask generation.
\label{fig}
\end{figure}

The predicted mask (Fig. 9c) demonstrates that the U-Net model is generally effective at identifying and segmenting the main blood vessels within the retinal image. The larger and more prominent vessels are clearly delineated, which is crucial for many clinical applications, such as diagnosing and monitoring retinal diseases.

\begin{figure}[htbp]
\centerline{\includegraphics[width=90mm,scale=0.7]{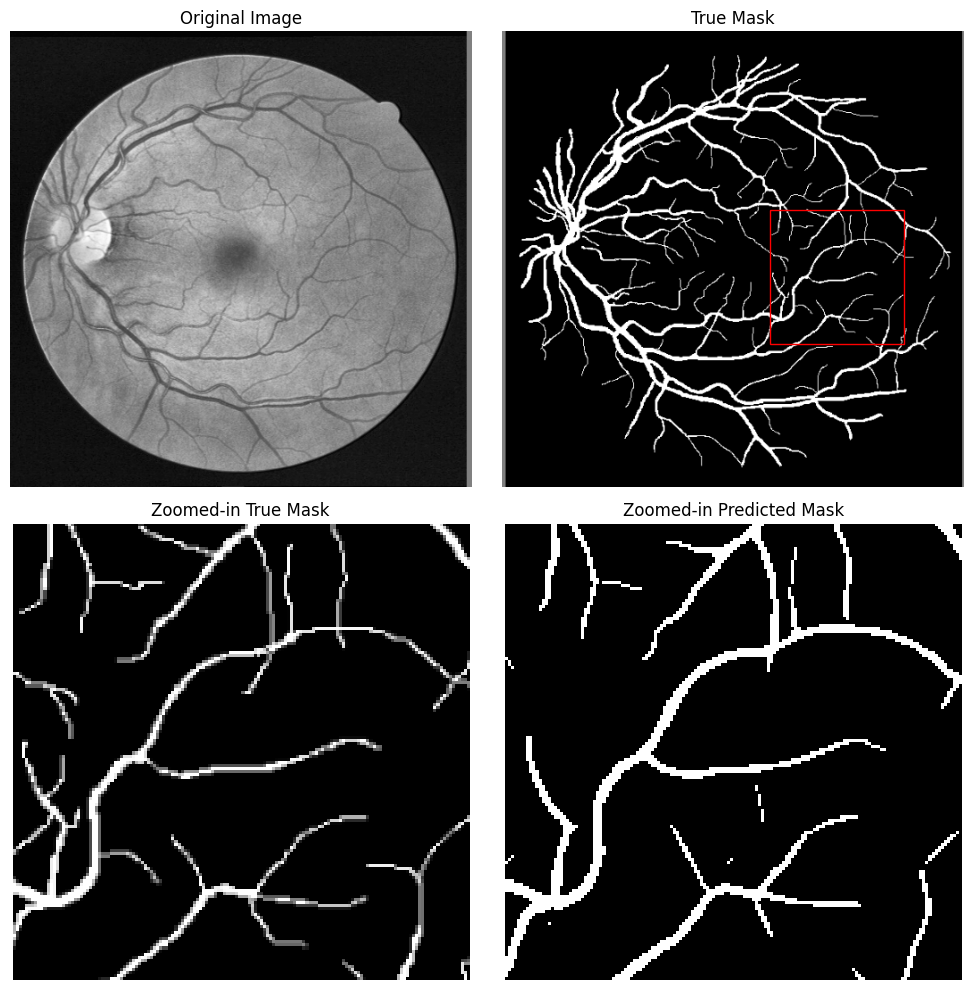}}
\caption{Comparison of magnified sections between true and predicted masks.}
\label{fig}
\end{figure}

However, despite the overall strong performance, the model exhibits limitations in detecting finer and more minute blood vessels. These smaller vessels are often essential for comprehensive retinal health assessments, yet they pose a challenge due to their subtle and less distinct appearance compared to larger vessels. As observed in Figure 10, several thin and intricate vessels are either partially detected or entirely missed. Additionally, predicted vessels sometimes appear thicker than their corresponding true  mask counterparts.  \\

We believe that the resolution (512x512) of the input images and the scale of the vessels significantly impacted the model's ability to detect fine details. The U-Net architecture, while powerful, would require higher resolution inputs or additional fine-tuning to capture the minutiae of retinal vasculature. However, this was not possible  in this study due to the amount of compute resources required for the processing of high resolution imagery. \\

We provide additional samples of predicted masks (Fig. 10) in order to demonstrate overall network efficacy.

\begin{figure}[htbp]
\centerline{\includegraphics[width=90mm,scale=0.7]{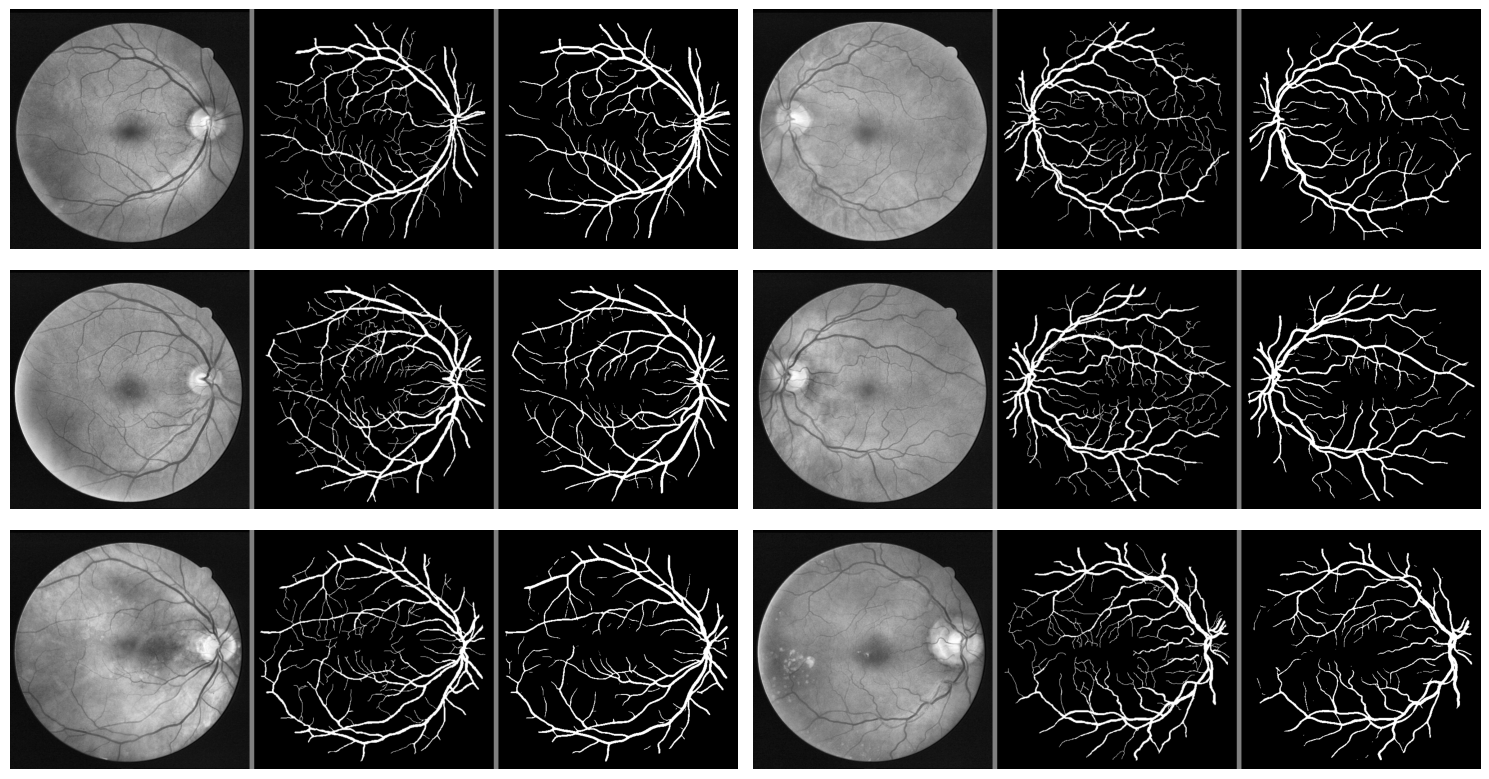}}
\caption{Testing images and their corresponding true and predicted masks.}
\label{fig}
\end{figure}
\subsection{Loss}
The model was trained using the DICE BCE loss function, which combines Dice loss and Binary Cross-Entropy loss. 
\begin{gather}
BCELoss(y, \hat{y}) = -(y\cdot log(\hat{y})+ (1-y) \cdot log(1-\hat{y}) \\
DiceLoss(y, \hat{y}) = 1 - \frac{2y\hat{y}}{y+\hat{y}+\epsilon} \\
DiceBCELoss(y,\hat{y}) = BCELoss(y, \hat{y}) + DiceLoss(y, \hat{y})
\end{gather}
Here, y is the actual value for a ground truth mask, and $\hat{y}$ is the value of a predicted mask. In Eq. 8, $\epsilon$ is added to ensure the function is not undefined in the edge case where $y + \hat{y} = 0$. We utilize this combination of functions in order to benefit from each of their advantages. BCE is powerful for pixel-wise classification, while Dice Loss is useful for improved boundary representation.  \\

The training and validation loss curves over 50 epochs (Fig. 12), show the model's learning dynamics.
\begin{figure}[htbp]
\centerline{\includegraphics[width=90mm,scale=0.7]{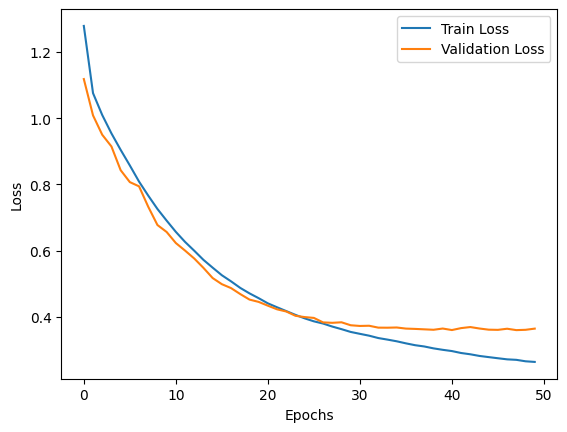}}
\caption{Training and validation loss curves of model.}
\label{fig}
\end{figure}
The training loss steadily decreases, indicating continuous learning and improvement of the model on the training data. The validation loss shows a similar decreasing trend, albeit with a slight plateau after epoch 40. This plateau suggests that the model begins to overfit.

\subsection{Metric Analysis}
The following table displays best metrics achieved by the model on the testing set. 
\begin{table}[htbp]
    \centering
    \begin{tabular}{|c|c|}
        \hline
        \textbf{Metric} & \textbf{Score} \\
        \hline
        Jaccard Index & 0.6766 \\
        \hline
        Precision & 0.8111 \\
        \hline
        Recall & 0.8051 \\
        \hline
        F1 Score & 0.8069 \\
        \hline
        Accuracy & 0.9650 \\
        \hline
        ROC-AUC & 0.8931 \\
        \hline
    \end{tabular}
    \label{tab:my_label}
\end{table}

We also provide formulas for each of the metrics in order to further discussion on model performance.

\subsubsection{Jaccard}
The Jaccard Index measures the overlap between the predicted segmentation and the ground truth. The score achieved by our model indicates it has a good ability to identify the vessel structures within the retinal images. Eq. 10 demonstrates how the metric is computed, where variables are the same as in Eq. 8.
\begin{align}
    Jaccard(y, \hat{y}) = \frac{y \cdot \hat{y}}{y + \hat{y} + \epsilon}
\end{align}
With a Jaccard score of 0.6766, the overlap between predicted and true masks is significant, but still appears to be missing a lot of information, likely due to the problems shown in Figure 10.
\subsubsection{Precision} In the context of this problem, the precision score achieved indicates that the majority of pixels identified as vessels by the model are indeed vessel pixels, which reduces the number of false positives.
\begin{align}
Precision = \frac{TP}{TP+FP}
\end{align}
\subsubsection{Recall} The recall of highlights the model’s proficiency in correctly identifying true vessel pixels. High recall is crucial in medical diagnostics to ensure that all vessel structures are detected.
\begin{align}
Recall = \frac{TP}{TP+FN}
\end{align}
\subsubsection{F1} The score achieved by our model demonstrates a strong balance between precision and recall. This metric is especially important in medical image segmentation, where both false positives and false negatives need to be minimized.
\begin{align}
F1 = \frac{2 \cdot Precision \cdot Recall}{Precision + Recall}
\end{align}
\subsubsection{Accuracy} Our very high accuracy shows that the model correctly classifies a large proportion of both vessel and non-vessel pixels across the dataset. However, it is not as useful of a metric as the other for this task, due to the fact that the dataset has a heavy pixel-wise class imbalance. That is, most of the pixels in each mask are black. 
\begin{align}
Accuracy = \frac{TP+TN}{TP+TN+FP+FN}
\end{align}
\subsubsection{ROC-AUC}
The Area Under the ROC Curve (AUC) provides an aggregate measure of performance across classification thresholds. The curve produced by our model (Fig 13.) demonstrates a sharp 'elbow' ROC curve (caused by the binary nature of masks) with an AUC value of 0.8931. We find this result satisfactory due to our limited time and compute resources in the development and training of the model. 
\begin{figure}[htbp]
\centerline{\includegraphics[width=90mm,scale=0.7]{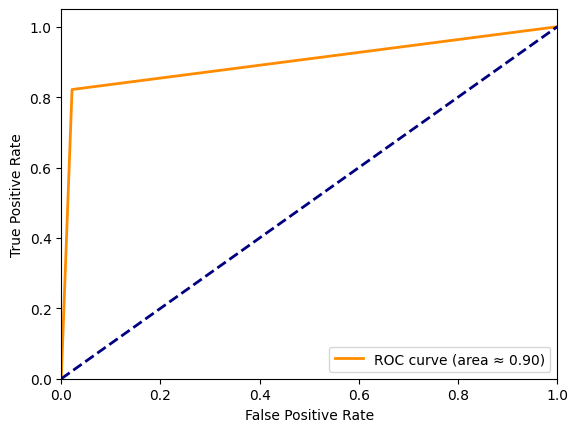}}
\caption{ROC curve.}
\label{fig}
\end{figure}

The following table compares the performance of the model developed in this study to the performance of other notable methods used on the DRIVE dataset. 

\begin{table}[htbp]
    \centering
    \begin{tabular}{|c|c|}
        \hline
        \textbf{Method} & \textbf{ROC-AUC score (DRIVE)} \\
        \hline
        \textbf{this study} & 0.8931\\
        \hline
        Azzopardi et al. [4] & 0.9614 \\
        \hline
        Soares et al [11] & 0.9614 \\
        \hline
        Osareh et al [5] & 0.9650 \\
        \hline
        Fraz et al. [8] & 0.9790 \\
        \hline
        Liskowski et al. [10] & 0.9650 \\
        \hline
    \end{tabular}
    \label{tab:my_label}
\end{table}

\section{Conclusion}
In this paper, we presented a U-Net based approach for the segmentation of retinal vessels, a critical task in the diagnosis and management of various retinal diseases. Our methodology leveraged the robust architecture of U-Net, optimized with a combined Dice and Binary Cross-Entropy (BCE) loss function to effectively manage class imbalance and improve segmentation accuracy. The metrics discussed in Section VI indicate the capability of the U-Net model to identify and segment retinal vessels. The high F1 score and precision values highlight the model's effectiveness in balancing between false positives and false negatives, which is crucial for reliable clinical applications. While the model is not as robust as other studies in the field (it fails to clearly identify minor vessels) it shows great potential in the speed of convergence and performance given very limited compute resources. 

Future research will focus on integrating certain improvements to enhance the model's performance further. Additionally, exploring other deep learning architectures and hybrid models could offer new insights and advancements in retinal vessel segmentation. Clinical validation of the model's predictions with larger and more diverse datasets will also be pursued to ensure its reliability and applicability in real-world settings.

Overall, our study demonstrates the efficacy of the standard U-Net model with added Batch Normalization and Dropout layers for retinal vessel segmentation, providing a foundation for its potential application in automated retinal disease diagnosis and monitoring.

\end{document}